\documentclass[prb,floatfix,twocolumn,amsmath,amssymb,showpacs]{revtex4}
\usepackage{graphicx}

\begin{document}

\title{Magnetic and transport properties of the one-dimensional
ferromagnetic Kondo lattice model with an impurity}

\author{S. Costamagna and J. A. Riera}
\affiliation{Instituto de F\'{\i}sica Rosario, Consejo Nacional de
Investigaciones Cient\'{\i}ficas y T\'ecnicas,\\
Universidad Nacional de Rosario, Rosario, Argentina
}

\date{\today}

\begin{abstract}
We have studied the ferromagnetic Kondo lattice model (FKLM) with an
Anderson impurity on finite chains with numerical techniques. We are
particularly interested in the metallic ferromagnetic phase of the
FKLM. This model could describe either a quantum dot coupled to
one-dimensional ferromagnetic leads made with manganites
or a substitutional transition metal impurity in a MnO chain. We 
determined the region in parameter space where the impurity is empty, 
half-filled or doubly-occupied and hence where it is magnetic or 
nonmagnetic. The most important result is that we found, for a wide 
range of impurity parameters and electron densities where the impurity 
is magnetic, a singlet phase located between two 
saturated ferromagnetic phases which correspond approximately to the 
empty and double-occupied impurity states. Transport properties 
behave in general as expected as a function of the impurity occupancy 
and they provide a test for a recently developed numerical approach to 
compute the conductance. The results obtained could be in principle 
reproduced experimentally in already existent related nanoscopic 
devices or in impurity doped MnO nanotubes.
\end{abstract}


\maketitle

\section{Introduction}
\label{intro}

Manganese oxides, such as La$_{1-x}$Ca$_{x}$MnO$_3$, commonly referred
to as manganites, have attracted an intensive theoretical and
experimental effort,\cite{lamno,dagorep} mainly due to their
property of colossal magnetoresistance\cite{colossal} and its
consequent applications to magnetic recording devices. General
applications of the ferromagnetic (FM) metallic phase of manganites
belong to the field of spintronics\cite{spinrev,ox-spin} where the
spin of the electrons is exploited in addition to its charge.

A simple spintronics device which is relevant for the present study
is formed by a quantum dot (QD),\cite{hanson} a nanometer-scale box,
connected to two FM leads.\cite{pasupathy} This device can
act as a spin valve\cite{martinek,gatorie} or a spin filter.
Ferromagnetic metals (Co, Pd-Ni) or diluted magnetic semiconductors,
such as GaMnAs, are employed as leads. Alternatively, manganites
are also used as FM leads in spintronics\cite{hueso,cottet} because
of its high polarization.

Manganites are usually described by the ferromagnetic Kondo lattice 
model (FKLM) in which the conduction sites represent the orbitals
$e_g$ and the localized spins the orbitals $t_{2g}$.\cite{lamno}
The QD will be described as a single Anderson impurity. The 
spin valve with manganites as leads corresponds then to a FKLM
with an Anderson impurity which is the model we will study in
the present work. Moreover, we will consider this model in a 
one-dimensional (1D) space. One should keep in mind also that
the magnetoresistance of manganites is usually applied in
multilayer heterostructures FM/M/FM or FM/I/FM (M: metal, I:
insulator) which can be considered as 1D systems in the 
direction perpendicular to the interface.\cite{hetero} 

This model, in addition to its 
application to a wide class of devices, could describe other
more conventional condensed matter systems such as a transition 
metal ion such as Cu replacing Mn in a manganese oxide 
chain.\cite{cromo} It is well-known that a single impurity could
lead to interesting and important local or short-range effects
in magnetic systems in low dimensions.\cite{martins} These 
effects can in turn modify the long-range physics of such systems
for a finite density of impurities.

The main purpose of this work is to search in the parameter space
of the model for phases in which the saturated FM is reduced to a 
partially polarized FM, or even to a nonmagnetic state, upon the 
introduction of an Anderson impurity. This problem would be
the analog for a ferromagnetic chain of the effect that causes a
magnetic impurity in a paramagnetic metallic chain, that is, the
paradigmatic Kondo effect.\cite{hewson}
We would like to emphasize that finite size effects due to our 
finite-cluster calculations could be relevant both to describe
mesoscopic devices and to capture local or short-range features
caused by an impurity in a Mn-O chain.
For completeness, since the model studied may be applied to 
electronic devices, we will compute the conductance through
the QD but clearly the study of transport properties is not the
main motivation for the present work. In any case, even though the
physics found for most of the parameter space corresponds to the
saturated FM phase and hence transport properties can be recovered
by a spinless fermion model, the impurity-FKLM is an interesting
testing ground for the quite recent numerical techniques we will
employ for this study.

\begin{figure}
\includegraphics[width=0.43\textwidth]{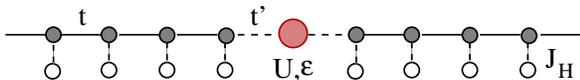}
\caption{(Color online) Picture of model (\ref{fklm-QD}). 
}
\label{fig1}
\end{figure}

\section{Model and methods}
\label{model}

Hence, in this article we will study a one-dimensional FKLM
with an Anderson impurity located in the center of the chain
(see Fig.~\ref{fig1}). Then, the model is defined by the 
Hamiltonian:
\begin{eqnarray}
{\cal H}_0 = &-& t_0 \sum_{i>0,i<-1,\sigma}
(c^{\dagger}_{i+1 \sigma} c_{i \sigma} + H.c. ) - 
J_H \sum_{l\neq 0} {\bf S}_{l}\cdot {\bf s}_{l} \nonumber\\
&-& t' \sum_{\sigma} (c^{\dagger}_{-1 \sigma} c_{0 \sigma} +
c^{\dagger}_{1 \sigma} c_{0 \sigma} + H.c. )   \nonumber\\
&+& U n_{0,\uparrow} n_{0,\downarrow} + \epsilon n_{0}
\label{fklm-QD}
\end{eqnarray}
\noindent
where the notation is standard. The Anderson impurity or ``QD", with 
parameters $U, \epsilon$, is located at site ``0" and is connected
to the rest of the system with a hopping $t'$. The ``leads"
($i\neq 0$) correspond to the FKLM with the Hund's rule exchange
coupling $J_H > 0$. $S_l$ is the spin operator for the localized
spin-1/2 orbital and $s_l$ is the one for the conduction
electron at site $l$ ($l\neq 0$). $t_0=1$ is adopted as unit
of energy, and we take $t'=0.4$ throughout. Model (\ref{fklm-QD})
will be termed ``QD-FKL" model.

The pure single-orbital 
FKLM or Kubo-Ohata model\cite{lamno} has been extensively studied,
particularly using numerical techniques\cite{rierahallberg} and its
phase diagram for various spatial dimensions and values $S$ of the
localized spins has been determined.\cite{dagotto} Even in the
simplest case of 1D and spin-1/2 localized spins, the model reproduces
qualitatively the main features of manganites. In the following we
will work in the metallic FM phase of the FKLM, typically, the
density of conduction electrons $n \leq 0.6$, $J_H=20$ (all coupling
constants are expressed in units of $t_0$). The on-site potential
$\epsilon$ and Coulomb repulsion $U$ are the main variables whose
effects
we want to study. In a heterostructure $\epsilon$ would be fixed by
chemistry but in a spin valve it would correspond to the gate
voltage which can be varied at will. In order to detect any departure
from the fully polarized FM state it is essential to work in
the subspace of total $S^z=0$ (1/2) for even (odd) number of
electrons.

We denote with $L$ the total length of the system including the
impurity site.
Open boundary conditions (OBC) were adopted in the lattices studied
except otherwise stated. Small clusters with $L$ up to 12 will
be studied using exact diagonalization (ED) with the Lanczos
algorithm. Larger clusters will be solved using density
matrix-renormalization group (DMRG).\cite{dmrgrev} For calculations
in the subspace of maximum total z-component of the spin, $S^z$,
i.e., saturated ferromagnetism, we used completely independent 
ED and DMRG codes for the spinless fermion model. The ED and DMRG
codes for the FKLM were thoroughly checked, in the first place 
by reproducing results in the literature. In the second place, by
comparing results obtained by both techniques in small clusters
and finally, by comparing results for large chains between FKLM
and the spinless fermion model in the case of maximum $S^z$. Here 
we would like to stress 
the fact that convergence to the ground state, both with ED and 
in the diagonalization of the superblock Hamiltonian at each 
iteration of DMRG is extremely slow. This is already known for 
ED studies of the Hubbard model with very large $U/t$ and small 
doping, that is in the proximity of the Nagaoka phase. 
DMRG studies for the Kondo lattice model (KLM), with both ferro- 
and antiferro- magnetic exchange coupling, have been in general
restricted to smaller chains than for the Hubbard model. In fact,
even in 1D, the DMRG treatment of KLM has the level of difficulty
of an interacting system on a two-leg ladder.
The convergence is even worst for the FKLM 
where previous studies have been limited to $L\approx 36$ with a
discarded weight of order $10^{-5}$.\cite{garcia}
Last but not least, the presence of impurities makes the convergence
more difficult particularly for DMRG. In our calculations, with a
retained number of $M\approx 400$, the truncation error is negligible
($\approx 10^{-14}$) for $L\approx 20$ in the regions close to
saturated ferromagnetism but it drops to $\approx 10^{-10}$ in the
nonmagnetic region. In the case of the spinless fermion model, for
$L\approx 20$ and $M\approx 400$, the precision in energy is at least
12 digits. In the parameter regions where total spin $S$ takes its
maximum possible value $S_{max}$, the energy in the $S^z=0$ subspace
reproduces the value obtained in the $S^z=S_{max}$ subspace using the
spinless fermion model within at least 9 digits. In any case, the
limited precision within DMRG depends essentially on the lack of
convergence in the diagonalization of the Hamiltonian. 

The conductance will be estimated by a numerical 
setup\cite{alhassanieh,schmitteckert} in which a small bias voltage
is applied to the left (L) and right (R) leads, with
$\Delta V=V_R-V_L$, ($V_R=-V_L$), at time $t=0$. The current
$J(t)$ induced by this voltage on each bond connecting the QD to
the leads is computed with the time evolution formalism both
within ED or DMRG.\cite{schollwock} This numerical setup is
equivalent to the systems which were treated analytically using
the Keldysh Green functions formalism.\cite{wingreen} These
results for out-of equilibrium, with interacting QD, contain
as particular cases the ones for noninteracting systems 
described by the Landauer formula.\cite{datta} These analytical
results, both for interacting and noninteracting QDs, were
recovered using this numerical setup and time-dependent
DMRG.\cite{cazalilla,alhassanieh,schmitteckert} The advantage of
this numerical procedure is that it can be extended with no
formal limitations to study the case of {\em interacting 
leads}\cite{qdhub} as will be done in the present work.

In principle, one could adopt as a
measure of the conductance the maximum of $J(t)/\Delta V$. It has
been shown that this recipe provides correct results for the 
conductance when the maximum of $J(t)$ corresponds to a ``plateau"
which appears in high-precision calculations using the 
``adaptive" time-dependent-DMRG on large clusters.\cite{alhassanieh}
In the following we adopt the less precise ``static"
algorithm\cite{cazalilla} which still gives qualitatively correct
results, particularly if relatively small clusters are considered,
but is much faster than the ``adaptive" scheme, thus allowing to 
explore a wider range of couplings and densities.
In the case of ED, the time evolution is exactly computed in the
full Hilbert space of the system. The time-evolution of the ground
state is given by
$|\Psi(t+\tau)\rangle = e^{-i{\cal H}\tau}|\Psi(t)\rangle$, where
${\cal H} = {\cal H}_0 + V_L N_L + V_R N_R$, $N_L$, $N_R$ are
the electron occupancies of the left and right leads respectively,
and $|\Psi(t=0)\rangle=|\Psi_0\rangle$,
${\cal H}_0 |\Psi_0\rangle = E_0 |\Psi_0\rangle$.
$|\Psi(t+\tau)\rangle $
was computed using the Krylov algorithm.\cite{manmana}
All the results reported below correspond to $\Delta V=0.01$,
$\tau=0.1$.

The dynamical impurity magnetic susceptibility and dynamical
magnetic structure factor (defined for convenience in 
Section~\ref{awaysymmetric}) are computed within ED and DMRG using
the standard continued fraction formalism. In the case of DMRG
we again choose the ``static" formulation which although less 
precise is enough to determine the presence or absence of a 
peak at the bottom of the spectrum.

We would like to stress that the most important results reported in
this article correspond to static properties, that is, ground state
energies and spin-spin correlations, where the precision of 
DMRG is maximal. For all quantities studied, the results obtained 
with ED are precise to precision machine.

\begin{figure}
\includegraphics[width=0.43\textwidth]{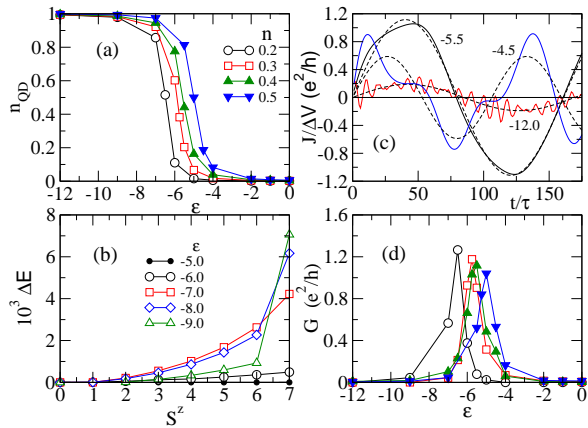}
\caption{(Color online) (a) QD occupancy as a function of $\epsilon$,
at several electron densities $n$. (b) $\Delta E=E(S^z)-E(0)$, 
versus $S^z$, $n=0.5$. (c) $J(t)/\Delta V$ for $n=0.4$,
$\epsilon=-4.5$, $-5.5$ and $-12$ (full curves).  Sinusoidal fits 
are shown with dashed lines. Curves for $\epsilon=-4.5$ and $-12$
have been multiplied by 2 and 20 respectively. (d) Conductance, as
a function of $\epsilon$, for the same electron densities as in (a). 
$L=10$, $J_H=20$, $t'=0.4$, $\epsilon=-U/2$.}
\label{fig2}
\end{figure}

\section{Results at the symmetric point}
\label{symmetric}

Let us start to analyze results for the QD-FKL model in the $L=10$
cluster obtained by ED. We consider in the first
place the case of the ``symmetric point", $\epsilon=-U/2$.
The symmetric point of an Anderson impurity is in principle the 
obvious place to look for a magnetic impurity placed in a 
noninteracting chain. However, this is not the case for the present 
model.
It can be
seen in Fig.~\ref{fig2}(a) that the QD or impurity occupancy, and
hence its actual magnetic or nonmagnetic character,
experiences a sharp crossover as a function of the on-site potential
of the impurity. For values of $\epsilon$ larger than a crossover
value $\epsilon^*$, $n_{QD}\approx 0$, and for values of
$\epsilon < \epsilon^*$, $n_{QD}\approx 1$. $\epsilon^*$ may be
defined as the value of $\epsilon$ at which $n_{QD}=0.5$. This
crossover can be understood by examining two variational states
in the atomic limit. One, with energy $E_1=-J_H n_e/4$, ($n_e$ is
the number of conduction electrons) where all electrons are located
on the leads and ferromagnetically aligned with the localized spins,
and the other with energy $E_2=-J_H (n_e-1)/4+\epsilon$,
where one electron has been moved from the leads to the QD. The
crossover between both variational states at
$\epsilon^*_{var}=-J_H/4$ is quite close to $\epsilon^*$
as shown in Fig.~\ref{fig2}(a). The dependence of $\epsilon^*$ 
with $n$ is mainly due to the kinetic energy which can be easily
computed within the spinless
fermion model to which the FKLM is reduced in the saturated FM state,
i.e. when total spin $S=S_{max}=S^z_{max}=n_e/2$.
In this case, neglecting the term with $t'$, $\epsilon^*_{spinless}$
is equal to the single particle energy of the top of the
band which increases with $n$ and is exactly zero at $n=0.5$.
The connection between both models implies that
$\epsilon_{spinless}=\epsilon+J_H/4$.

It is
important to notice at this point that although the pure system
is in the saturated FM state for the densities studied, for some
impurity parameters the impurity may drive the system into
partially polarized FM states with total spin $S<S_{max}$. 
In fact, as shown in  Fig.~\ref{fig2}(b) for $n=0.5$, there is a
region of $\epsilon$, close to $\epsilon^*$, where $S=0$ ($1/2$)
for even (odd) $n_e$. We have observed this nonmagnetic state for other
chain lengths and densities. Although the difference in energy
between states with different $S^z$ is very small, these results
strongly suggest that there are regions in parameter space where the
impurity causes a breakdown of the fully saturated FM state.
This possibility will be thoroughly examined in the next
Section.

Let us discuss in detail how the conductance $G$ is determined.
In Fig.~\ref{fig2}(c), it is shown
$J(t)/\Delta V$ ($J(t)$ is the average of the current on the two
bonds connecting the QD to the leads) which presents the typical
oscillatory behavior.
This oscillatory behavior follows from the expansion of 
$e^{-i{\cal H}\tau}$ in eigenvectors of ${\cal H}$, which for small
$\Delta V$ are adiabatically related to those of ${\cal H}_0$. We
would like to emphasize that results depicted in Fig.~\ref{fig2}(c)
are exact, i.e., no truncation of the Hilbert space was performed.
Then, we fit each curve by a sinusoidal
and we adopt $G$ as the amplitude of this sinusoidal. In this
small cluster, but also for $L=20$, a single sinusoidal gives a 
reasonable fitting of $J(t)/\Delta V$ for most of the cases studied,
particularly near $\epsilon^*$. Although this procedure is not very
precise, it gives qualitatively correct results as we discuss in 
the following.

Results for the conductance are shown in Fig.~\ref{fig2}(d) for
various densities as a function of $\epsilon$. $G$ is only different
from zero at the crossover between the region of empty QD
($\epsilon>>\epsilon^*$) and the region of half-filled QD
($\epsilon<<\epsilon^*$) and it has a sharp peak at $\epsilon^*$
with a width approximately equal to the bandwidth of a tight-binding
model on the leads, $4 t_0$. This behavior is what one would expect
for the spinless fermion model.  The determination
of the variation of the maximum conductance with density $n$,
which would require calculations on a finer mesh in $\epsilon$,
is out of the scope of the present study.

\begin{figure}
\includegraphics[width=0.43\textwidth]{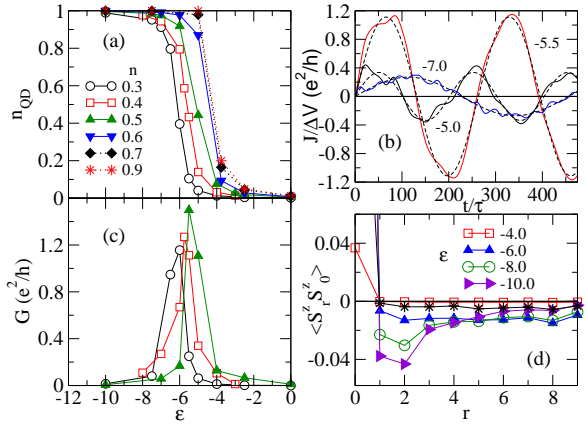}
\caption{(Color online) (a) QD occupancy and (c) conductance, as a
function of $\epsilon$, for several electron densities $n$ indicated
on the plot. (b) $J(t)/\Delta V$ for $n=0.4$, $\epsilon=-5$, $-5.5$ 
and $-7$ (full curves). Fits to a sine function are shown with dashed
lines. Results for $L=20$, $J_H=20$, $t'=0.4$, $\epsilon=-U/2$. 
(d) Spin-spin correlations along the conduction chain, for the pure
FKLM (stars) and in the presence of an impurity for various values of
$\epsilon$ indicated on the plot, $L=19$, $n=0.421$. The reference
site is located at the center of the chain and
the normalization $\langle S^z_0 S^z_0 \rangle =1$ was adopted.}
\label{fig3}
\end{figure}

Let us now discuss results for $L=20$, obtained with DMRG, also
at the symmetric point. In Fig.~\ref{fig3}(a), it can be seen that,
as for $L=10$, the impurity occupancy $n_{QD}$ experiences a
sudden change as a function of the on-site potential $\epsilon$.
This crossover is located approximately at $\epsilon^*=-J_H/4$ as
argued before, even inside the incommensurate (IC) phase (but not
strictly at $n=1$), which is expected since the variational states
are independent of the underlying FM or IC order.
As in the $L=10$ cluster, the location of this crossover shifts to
larger values of $\epsilon$ as the density is increased. At the IC-FM
crossover for $J_H=20$, $n\approx 0.55$, $\epsilon^*$ experiences
a somewhat larger increase and then it remains relatively
unchanged up to half-filling. In the IC region of course the kinetic
energy is no longer approximated by the spinless fermion model.
In fact, it is easy to realize that the kinetic energy
versus $\epsilon$ follows an {\em opposite} behavior as the
IC-FM border is crossed.

The computation of the conductance follows the steps previously 
outlined. In Fig.~\ref{fig3}(b), $J(t)/\Delta V$ is shown for
$n=0.4$ and several values of $\epsilon$. It can be seen that
in spite of the approximate nature of the computation of the 
time evolution in a truncated Hilbert space, $J(t)$ is clearly
well fitted by a single sinusoidal, particularly close
to $\epsilon^*$, and these oscillations have similar behaviors
 as a function of $\epsilon^*$ as earlier for the $L=10$ chain.
In Fig.~\ref{fig3}(c) we show the resulting $G$ as a function of
$\epsilon$ and for several densities. As for the
smaller lattice $L=10$, the conductance is different from zero only
for $\epsilon \approx \epsilon^*$, with a peak at
$\epsilon^*$.\cite{note2}

In spite of the extremely slow convergence in DMRG calculations, 
by keeping 450 states we were able to find that
$E(S^z=0)< E(S^z=S_{max})$ for $\epsilon=-8$, thus suggesting
for $L=19$ a similar behavior to that shown in Fig.~\ref{fig2}(c)
for $L=10$. Further indications of this behavior can be obtained
by examining the $z$-component of the spin-spin correlations
$\langle S^z_j S^z_0\rangle$, where the reference site ``0" is the
center of the chain and $j$ labels conduction sites. Due to the large
value of $J_H$, the correlations between the impurity site and the
localized spins have the same qualitative behavior so in this and in 
the following section we will only consider the correlations between 
the impurity and the conduction sites. These correlations, 
shown in Fig.~\ref{fig3}(d) for $n=0.421$, clearly depart from the
correlations in the pure system as  $|\epsilon|$ is increased.
This behavior indicates that the
ground state computed by DMRG is a mixture of states very close in
energy which depart from the saturated FM state. Then, this behavior
of $\langle S^z_j S^z_0\rangle$ with $\epsilon$, which follows the same 
trend as the one for the $L=10$ chain, suggests that also for the $L=19$
chain the ground state $S < S_{max}$. Notice also that in this
low-$S$ region, $\langle S^z_j S^z_0\rangle$ do not show any trace of
antiferromagnetic order.

\begin{figure}
\includegraphics[width=0.4\textwidth]{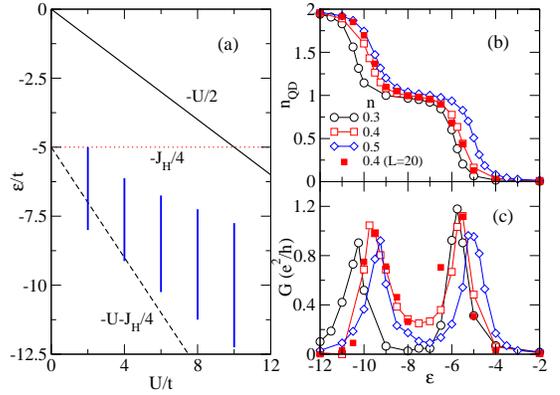}
\caption{(Color online) (a) Variational states for model
Eq.~(\ref{fklm-QD}) in the $U$-$\epsilon$ plane. Vertical thick lines
indicate approximately the region with minimum $S$ for some values of
$U$, $n=0.4$, discussed in the text. (b) QD occupancy and (c)
conductance as a function of $\epsilon$, for $U=4$, and various electron
fillings $n$.  Results for $L=10$ chain except otherwise stated.
}
\label{fig4}
\end{figure}

\section{Breakdown of the ferromagnetic state away from the symmetric
point}
\label{awaysymmetric}

Let us discuss the consequences of these results for devices where
the leads are made with manganites. Since $J_H/t_0$ in manganites
has been estimated of the order or larger than 10,\cite{lamno,dagorep}
then a value of $U^* = -2 \epsilon^* \geq 5$ in the QD (see 
Fig.~\ref{fig4}(a)) would be required for $n_{QD}\sim 1$.
This value of $U$ is somewhat larger than the one in materials
employed in the QD, such as semiconductors or carbon nanotubes.
It is necessary then that the device could be operated away from the
symmetric point. More importantly, a smaller $U$ could imply a larger
effective coupling between the impurity and the conduction sites, 
assuming that to lowest approximation the relation 
$J_{eff} \approx t'^2/U$ is still valid for the present model for
$U > t'$. In support of this hypothesis, we have observed that the
spin-spin correlation between the impurity and its nearest neighbor
site becomes more negative with decreasing $U$ at fixed $\epsilon$.
Then, by working with a larger $J_{eff}$ we could expect to be more
able to detect the presence of the nonmagnetic phase which was  
suggested by the results found in the previous section.
For these reasons, in the following we adopt a moderate value of
the on-site Coulomb repulsion, $U=4$, and we study the properties
of the model for variable $\epsilon$, i.e. following a vertical line
in Fig.~\ref{fig4}(a).

\begin{figure}
\includegraphics[width=0.43\textwidth]{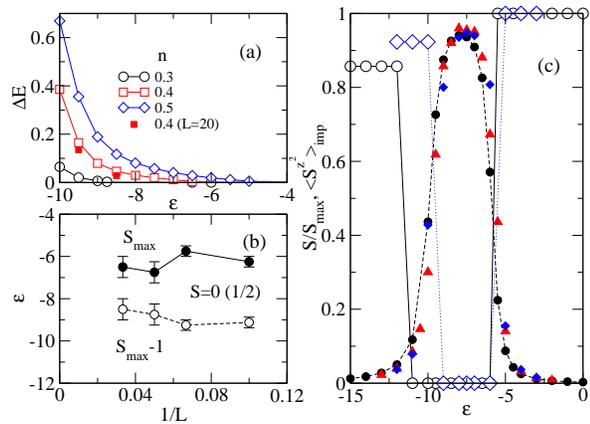}
\caption{(Color online) (a) 
$\Delta E =E(S^z_{max})-E(S^z_{min})$, as a function of $\epsilon$,
for $U=4$, and various electron fillings $n$. Results for $L=10$
chain except otherwise stated. (b) Upper (solid line) and lower
(dashed line) boundaries of the region where the ground state
$S = 0 (1/2)$, for $n=0.4$, as a function of $L$, for $U=4$, and
$n=0.4$. (c) $<S^{z^2}>$ (normalized to 1) at the impurity site (full
symbols) and total spin normalized to $S_{max}$ (open symbols), as a
function of
$\epsilon$, for $U=4$. Results for $L=11$, $n=0.364$ (circles),
$L=19$, $n=0.421$ (diamonds), and $L=20$, $n=0.4$ (triangles).}
\label{fig5}
\end{figure}

The electron occupancy at the QD, shown in Fig.~\ref{fig4}(b) for
$L=10$, presents now three regions where $n_{QD}$ is approximately
0, 1, and 2 as $\epsilon$ decreases.
The crossovers among these regions are located near the
variational estimates $\epsilon^*_{0,1} \approx -J_H/4$, and
$\epsilon^*_{1,2} \approx -J_H/4 -U$ which are shown in 
Fig.~\ref{fig4}(a).
As it can be seen in Fig.~\ref{fig4}(c), the conductance $G$
consistently with the results shown in Figs. ~\ref{fig2}(d) and
~\ref{fig3}(c) at the symmetric point, presents sharp peaks at the
crossovers between regions with different $n_{QD}$, 
i.e. when $n_{QD}\approx 0.5$ or $1.5$. Notice that $G$ has a
larger value between the peaks for $n=0.4$ compared with the one
for $n=0.3$, which is essentially zero.

Figure~\ref{fig5} contains the most important result of our work. In
Fig.~\ref{fig5}(a) we plot $\Delta E =E(S^z_{max})-E(S^z_{min})$,
where $S_{min}=S^z_{min}=0$ ($1/2$) for even (odd) $n_e$, as a
function of $\epsilon$ in the $L=10$ chain. It can be clearly seen
that the ground state $S$ is smaller than $S_{max}$ for the various
densities considered for $\epsilon  \leq \epsilon^*_{0,1}$.  In 
this case, $\Delta E$ is quite large and we were able to obtain for
$L=20$ results very close to those for $L=10$, as shown for $n=0.4$,
suggesting that this feature is at least not an artifact of this
small chain.
More interesting is the fact that inside the $n_{QD}=1$ region there
is an interval in $\epsilon$, which depends on the density, where 
$S=S_{min}$, as shown in Fig.~\ref{fig5}(b) for $n=0.4$ as a 
function of $1/L$. This state appears between two saturated FM
states, $S=S_{max}-1$ below the lower boundary line 
because of the double occupancy of the QD, and $S=S_{max}$ above
the upper boundary line where the QD is empty. The error bars
quoted in the plot correspond to the grid adopted in the $\epsilon$
axis. For the largest chain studied, $L=30$, we have only computed
the energies in the $S^z=S_{max}$, $S_{max}-1$, $S_{max}-2$ and
$S_{max}-3$ subspaces. Decreasing precision as $L$ increases
prevents us to take a finer grid close to the crossover between 
different regions and hence error bars are larger. An extrapolation
to the bulk limit would not be reliable with these error bars.
For $L=10$, $n=0.3$, the
region in $\epsilon$ where $S=1/2$ shrinks to $[-9.3,-8.6]$. It
is tempting to relate this smaller interval for $n=0.3$ with respect
to the one for $n=0.4$ with the behavior of $G$ noticed above but
further analysis would be needed to confirm this possibility. In 
Fig.~\ref{fig5}(c) we show that the presence of a state with
$S=S_{min}$ corresponds to a strong magnetic character of the
impurity with $<S^{z^2}>$ approaching its maximum value of one 
(according to the normalization adopted). This feature
has been observed for all lattices, densities, and values of $U$
studied. The intervals in $\epsilon$ where $S=S_{min}$ for various
values of $U$, $n=0.4$, $L=10$ are shown in Fig.~\ref{fig4}(a) 
with thick lines.

\begin{figure}
\includegraphics[width=0.43\textwidth]{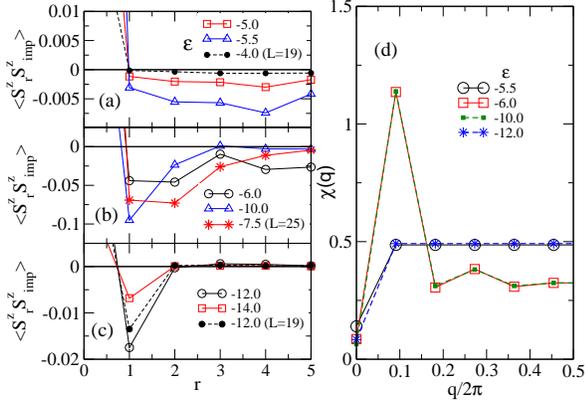}
\caption{(Color online) Spin-spin correlations from the impurity site
along the conduction chain for several values of
$\epsilon$ in the regions where 
(a) $S=S_{max}$, (b) $S=0$, and (c) $S=S_{max}-1$. The normalization
$\langle S^z_0 S^z_0 \rangle =1$ was adopted.
(d) Static structure factor along the conduction chain for various 
values of $\epsilon$. Results for $L=11$ chain, PBC, $U=4$, $n=0.364$,
except otherwise stated. Results for $L=19$ correspond to $n=0.421$
and for $L= 25$ $n=0.4$.
}
\label{fig6}
\end{figure}

The behavior of the magnetic character of the impurity affects in turn
all the magnetic properties in the system. Let us first examine the
spin-spin correlations between the impurity site and the remaining sites
along the conduction chain. Results for the $L=11$ chain with periodic
boundary conditions (PBC), $n=0.364$, are shown in
Fig.~\ref{fig6}(a), (b) and (c) in the three regions where the spin of
the ground state is $S_{max}$, 0 and $S_{max}-1$, respectively.
We also included in this figure results for $L=19$, $n=0.421$, and $L=25$, 
$n=0.4$, OBC, which show the same behavior. It should
be noticed that in addition to the expected larger magnitude of these 
correlations in the $S=0$ region with respect to the ones in the other two
regions, there are also qualitative changes. To detect these qualitative
differences we have computed the static structure factor along the
conduction chain:
\begin{eqnarray}
\chi(q) = \frac{1}{L} \sum_{l,j} \langle S^z_l S^z_j\rangle
e^{i q (l-j)}
\label{formfact}
\end{eqnarray}
\noindent
where $j,l$ label the conduction chain sites, and 
$q=(2\pi/L)n$, $n=0,\ldots,L-1$ . Results where site $j$
is restricted to the impurity site, that is just the Fourier transform
of the correlations shown in Fig.~\ref{fig6}(a), (b) and (c), are 
essentially the same as those obtained using Eq.~(\ref{formfact}).
As it can be seen in Fig.~\ref{fig6}(d), $\chi(q)$ has the typical
shape of a FM order in the subspace of total $S^z=0$ in the regions
with $S=S_{max}$ and $S_{max}-1$, while it presents a peak at the
smallest nonzero momentum in the region $S=0$. As observed earlier,
in Fig.~\ref{fig3}(d), there are no traces of AF order in this region.

Further information about changes in magnetic properties caused by
the impurity can be obtained by looking at the dynamical 
impurity susceptibility defined as:
\begin{eqnarray}
S_{imp}(\omega) = \sum_{n} |\langle \Psi_n | S^z_{imp} |
\Psi_0 \rangle |^2 \delta(\omega -(E_n-E_0))
\label{dyn-rs}
\end{eqnarray}
\noindent
where the notation is standard.
Since the contribution from the remaining sites on the conduction
chain is negligible, $S_{imp}(\omega)$ is essentially equal to the
total dynamical susceptibility once the contribution from localized
spins has been subtracted. In all results below, the peaks have been
broadened with a width $\delta=0.1$.

\begin{figure}
\includegraphics[width=0.43\textwidth]{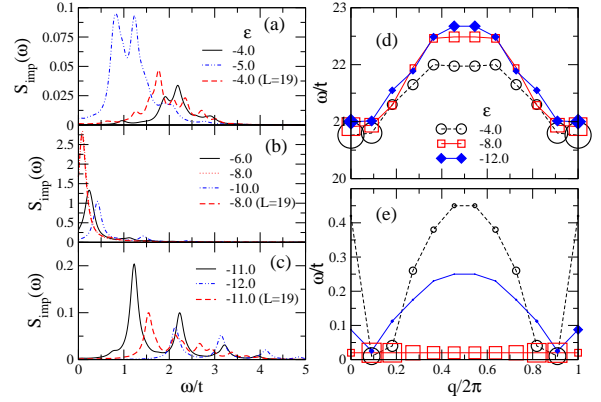}
\caption{(Color online) Dynamical impurity susceptibility
$S_{imp}(\omega)$ vs $\omega$ for $U=4$, $n=0.4$, and values of
$\epsilon$ in the region where (a) $S=S_{max}$, (b) $S=0$, and (c)
$S=S_{max}-1$.
Energies of the dominant peaks in the dynamical structure factor
$\chi(q,\omega)$, (d) in the high- and (e) in the low-$\omega$
regions as defined in the text for various values of $\epsilon$.
The size of the symbols is proportional to the intensity of each
peak. Results for $L=11$ chain except otherwise stated.
}
\label{fig7}
\end{figure}

In Fig.~\ref{fig7}(a), (b) and (c) we show $S_{imp}(\omega)$ in the 
low-$\omega$ part of the spectrum for various values of $\epsilon$
in the regions $S=S_{max}$, 0 and  $S_{max}-1$ respectively, for 
$L=11$, $n=0.4$ and $L=19$, $n=0.421$. Again, in addition to an
expected difference in amplitude, a clear qualitatively different
behavior is
noticeable. In this part of the spectrum $S_{imp}(\omega)$ presents
small peaks at finite $\omega$ in the $S=S_{max}$ and $S_{max}-1$
regions while in the $S=0$ region, $S_{imp}(\omega)$ shows a strong
peak close to $\omega=0$. Results for $L=25$, $n=0.4$, $\epsilon=-8$ 
are indistinguishable from those for $L=19$.

The distinct behavior caused by the impurity can also be detected
by looking at the dynamical structure factor on the conduction chain
which is defined as:
\begin{eqnarray}
\chi(q,\omega) = \sum_{n} |\langle \Psi_n | S^z_{q} |
\Psi_0 \rangle |^2 \delta(\omega -(E_n-E_0))
\label{dyn-mom}
\end{eqnarray}
\noindent
where $S^z_{q}=(1/L)\sum_j S^z_j \exp{(ijq)}$, and the sum extends 
over the conduction chain.
Figure~\ref{fig7}(d) and (e) show the energy and intensity of the 
centroid of the main peaks in the high- and low-$\omega$ parts of
the spectrum for values of  $\epsilon$ in the three regions of
total spin $S$ above discussed. In the $S=S_{max}$ and $S_{max}-1$ 
regions the peaks with largest weight are those with energy
close to $J_H$ which corresponds to the magnon excitation. This
behavior is strikingly different to the one in the $S=0$ region
where the most weighted peak form a dispersionless band at the
bottom of the spectrum. This band is reminiscent of the one found
in gapped spin systems upon doping with nonmagnetic 
impurities.\cite{martins} In the present case, these low-energy
peaks may correspond to magnetic excitations living in a ``cloud"
surrounding the impurity that can be observed in real-space in
Fig.~\ref{fig6}(b).

\section{Discussion and conclusions}
\label{conclusions}

We have applied both well-established and recently developed numerical
approaches to study the FKLM with an Anderson impurity in the
FM phase on finite chains. We found that the magnetic or nonmagnetic
character of the impurity is determined by a relationship between
the impurity parameters and the Hund's rule exchange coupling of
the manganite. As expected, transport occurs at the crossovers
between the empty, half-filled or filled QD regions.

The most important result of this article is the presence of an
intermediate singlet phase between the fully saturated phases with
$S_{max}$ (empty impurity) or $S_{max}-1$ (doubly occupied impurity).
This problem has some resemblance with the problem of the existence
of an  intermediate phase between two ordered states in the frustrated
Heisenberg model on the square lattice. This
intermediate phase had been predicted by ED studies on small
clusters.\cite{j1j2inter} The alternative view was 
a first order quantum phase transition between both ordered phases.
Only recently was this controversy being settled favouring the
existence of the intermediate state\cite{j1j2fin} but the nature of
this phase is still a subject of active research.\cite{j1j2nuevo}
Of course the physics involved in both problems is completely
different, but by analogy, in our problem, we could consider the
possibility of a first-order transition between two FM states instead
of the proposed intermediate nonmagnetic phase.  By comparing
the models involved in these two problems, FKLM and frustrated
Heisenberg model, the former is much more difficult to analyze
than the latter by numerical techniques since the size of the
Hilbert space is much larger for a given cluster and moreover taking
into account the convergence problems discussed in 
Section~\ref{model}. These difficulties prevent us to perform
an extrapolation in order to decide if this intermediate phase
really exists in the bulk limit or if it is just how the
transition between the $S_{max}$ and $S_{max}-1$ phases
manifests in finite systems.

In any case, the possibility of the
existence of this intermediate state is in principle interesting 
and important and it deserves further study. In this sense,
there are three issues that should be considered. The first one
is that this intermediate phase could be stabilized by some
modifications of the model, for example by including the
Heisenberg interaction between localized spins, or by replacing the
spin-1/2 localized spins by higher spin ones which, in addition, are
also more realistic for manganites. The second issue is that even if
this intermediate phase has a finite range around an impurity,
a {\em finite density} of impurities could
lead to a macroscopic feature. The situation here is analog to
the presence of nonmagnetic impurities in the above mentioned
gapped systems.\cite{martins} In these systems a single impurity
attracts locally a spinon to the impurity and a finite density
of impurities drives the system to a long-range AF order. These
two issues are currently under study.\cite{largo} The third issue
we would like to consider is related to the relevance of the
present model to devices with {\em finite} dimensions.
In these mesoscopic systems, as discussed
in introductory textbooks,\cite{datta} due to its finite size,
many physical properties are different to that found in bulk
systems. It is then relevant for these devices to capture 
short-range effects.

Finally, we would like to provide a qualitative scenario to help
understanding this nonmagnetic state. Let us assume that the system
is in a low $S^z$ state. In the region where the impurity is empty
or double occupied, both ``leads" are relatively disconnected and
each one would have a ferromagnetic state with spins polarized in 
one direction and the other with spins polarized in the opposite
direction. Of course this state is degenerate with the one with
reversed polarizations. Now, when the impurity is singly occupied,
not only it would have a definite magnetic character but it would
allow the crossing of one electron from one lead to the other where
it would have then a ``wrong" spin. This kind of magnons would then
decrease the total energy both by increasing the kinetic energy and by
decreasing the magnetic energy due to an effective AF interaction
with the impurity. Of course, this gain in energy would not occur
for the fully polarized system so it would drive the system to
lower $S$ and presumably to $S_{min}$.\cite{note3} It is interesting
to notice that this scenario would then imply an enhancement of
transport in the system which could be relevant for the devices 
mentioned in the Introduction.

In summary, we present a prediction on the magnetic state of
the FKLM doped with a magnetic impurity. This prediction could
be experimentally verified on Cu-doped manganite nanotubes. These
results could also be in principle reproduced experimentally on
spin valves where manganites are used as ferromagnetic leads.
We hope the present results will encourage theoretical studies to 
further characterize this proposed intermediate phase and to
explore its presence in more realistic models for manganites.

\acknowledgments
We thank E. Dagotto, A. Dobry, C. J. Gazza, M. E. Torio, and S.
Yunoki for useful discussions.


\begin{thebibliography}{}
\bibitem{note2} For chains with $L$ odd, we obtain that the conductance
     maximum is located at values of $\epsilon$ lower (higher) than
     $\epsilon^*$ for even (odd) number of conduction electrons. However,
     the results for $L$ odd converge to those of $L$ even as the chain
     length is increased.
\bibitem{note3} Preliminary calculations show that the magnetic energy
     decreases strongly than the increase of kinetic energy as $S^z$ is
     reduced.
\end{thebibliography}
\end{document}